\documentclass[final,numberedheadings]{aipproc}

\layoutstyle{6x9}
\usepackage{bm}

\newcommand{\ff}[1]{{\boldsymbol #1}}
\newcommand{\tr}{\mbox{tr}}
\newcommand{\Tr}{\mbox{Tr}}

\begin{document}

\title{Systematics of approximations constructed from dynamical variational principles}

\classification{71.10.-w, 71.15.-m}

				

\keywords{Variational principles, lattice fermion models, dynamical mean-field theory, 
          cluster approaches, Hubbard model}

\author{Michael Potthoff}{address={Institut f\"ur Theoretische Physik und Astrophysik, Universit\"at
W\"urzburg, Germany}
}

%

\begin{abstract}
The systematics of different approximations within the self-energy-functional 
theory (SFT) is discussed for fermionic lattice models with local 
interactions.
In the context of the SFT, an approximation is essentially given by 
specifying a reference system with the same interaction but a modified 
non-interacting
part of the Hamiltonian which leads to a partial decoupling of degrees of 
freedom.
The reference system defines a space of trial self-energies on which an
optimization of the grand potential as a functional of the self-energy
$\Omega[\ff \Sigma]$ is performed.
As a stationary point is not a minimum in general and does not provide a
bound for the exact grand potential, however, it is {\em a priori} unclear 
how to judge on the relative quality of two different approximations.
By analyzing the Euler equation of the SFT variational principle,
it is shown that a stationary point of the functional on a subspace
given by a reference system composed of decoupled subsystems is also a
stationary point in case of the coupled reference system.
On this basis a strategy is suggested which generates a sequence of 
systematically improving approximations.
The discussion is actually relevant for {\em any} variational approach 
that is not based on wave functions and the Rayleigh-Ritz principle.
\end{abstract}

\maketitle

\section{Introduction}
\label{sec:intro}

Lattice models of correlated electrons such as the single-band Hubbard model 
\cite{Hub63,Gut63,Kan63} represent one of the central issues in solid-state
theory. 
One reason for this strong interest is that the Hubbard model is one of the
simplest but non-trivial models that allow for a benchmarking of new 
theoretical concepts. 
In the recent years, dynamical cluster approaches to the Hubbard model and 
its variants have become more and more popular \cite{MJPH04}.
Contrary to techniques that are based on the Ritz variational principle and
on the optimization of wave functions, {\em dynamical} cluster concepts not 
only give information on the static thermodynamic properties of a system 
but also on the elementary single-particle excitations. 
These approaches can be divided into two groups: 
(i) cluster extensions of the dynamical mean-field theory (DMFT) 
\cite{GKKR96,MV89} and (ii) variational cluster extensions of the simple 
Hubbard-I approximation \cite{Hub63}.

The DMFT can be understood as a mean-field theory which neglects spatial
correlations but which fully takes into account temporal fluctuations.
This is reflected in the DMFT approximation for the self-energy 
$\Sigma_{ij}(\omega)$ which is local in the site indices 
$\Sigma_{ij}(\omega) = \delta_{ij} \Sigma(\omega)$ but shows up a non-trivial 
and in general strong dependence on the excitation energy $\omega$.
Spatial correlations are systematically restored in the
dynamical cluster approximation (DCA) \cite{HTZ+98,MJPH04} or in
the cellular DMFT (C-DMFT) \cite{KSPB01,LK00}.
Contrary to the original DMFT, where the self-energy is generated from a
model where a single correlated site is embedded into a continuous 
non-interacting medium (``bath''), the cluster extensions employ more 
complicated reference systems where the single-site impurity is replaced 
by a cluster of $L_{\rm c} > 1$ correlated sites.
This generates self-energies with off-site elements.
The bath parameters are determined by a so-called self-consistency equation 
which relates the reference (impurity/cluster) model with the original 
(Hubbard) model.

The Hubbard-I approximation represents a very simple approximation scheme
which originally was constructed \cite{Hub63} by a more or less {\em ad hoc}
decoupling of the equations of motion for the one-particle Green's function. 
Equivalently, however, it can be understood as a scheme which approximates
the self-energy of the original Hubbard model by the self-energy of an 
atomic model consisting of a single correlated site ($L_{\rm c}=1$).
From this perspective, a cluster generalization is straightforward and 
yields the cluster-perturbation theory (CPT) \cite{GV93,SPPL00}.
The CPT can also be considered as the first non-trivial order in a systematic 
expansion in powers of the inter-cluster hopping parameters.
The usual CPT uses a cluster of finite size $L_{\rm c} > 1$ which is cut
out of the original lattice to generate the approximate self-energy.
The main idea of the variational CPT (V-CPT) \cite{PAD03,DAH+04} is to optimize
the self-energy by varying the parameters of the cluster.
This is reminiscent of the optimization of the self-energy by varying the 
bath parameters in the cluster extensions of the DMFT.
For the construction of a thermodynamically consistent approximation the
variational aspect is essential \cite{AAPH05}.
It is therefore reasonable to call this a variational cluster approach
(VCA).

\begin{figure}[t]
\centerline{\includegraphics[width=80mm]{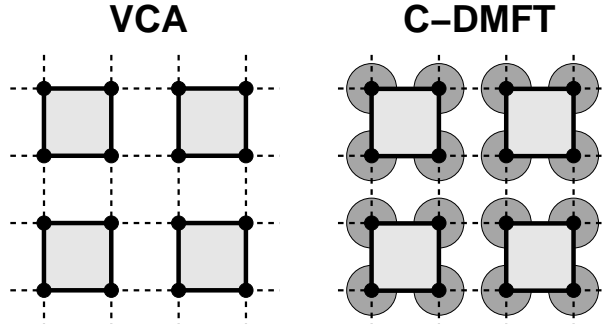}}
\caption{
Illustration of two reference systems $H'$ generating two different approximations:
The variational cluster approximation (VCA) and the cellular dynamical mean-field theory
(C-DMFT). The reference systems consist of $L_{\rm c} > 1$ correlated sites (non-zero 
Hubbard-$U$) per cluster. The intercluster hopping (dashed lines) is switched off. 
In case of the C-DMFT an uncorrelated, continuous bath is attached to each of the 
original sites in addition.
For both, the VCA and the C-DMFT, all one-particle parameters of $H'$ are considered
as variational parameters.
}
\label{fig:apps}
\end{figure}

Both types of approximations, (i) and (ii), can be obtained within a general
framework which is known as the self-energy-functional theory (SFT) 
\cite{Pot03a,Pot05}.
The SFT is based on a variational principle $\delta \Omega_{\ff t,\ff U}[\ff \Sigma] = 0$
\cite{Pot04} which goes back to the original ideas of Luttinger, Ward, Baym 
and Kadanoff in the sixties \cite{LW60,BK61} and which provides a very general
framework to construct dynamical approximations.
Let the original (Hubbard-type) model with one-particle and interaction parameters
$\ff t$ and $\ff U$ be given on a lattice consisting of $L$ sites (with $L \to \infty$).
Consider then a partitioning of the lattice into disconnected clusters with a finite 
number $L_{\rm c}$ of correlated sites (and possibly also a number of $L_{\rm b}$ additional 
uncorrelated bath sites attached to each of the correlated sites).
The model on the truncated lattice (the ``reference system'') is therefore given by 
modified one-particle parameters $\ff t'$ and serves to define trial self-energies 
$\ff \Sigma_{\ff t',\ff U}$ for the variational principle. 
The trial self-energy is varied by varying the one-particle parameters of the reference 
system $\ff t'$. 
In this way one can search for a stationary point of the self-energy functional 
on the restricted space of self-energies defined by a simpler reference system:
\begin{equation}
\frac{\partial}{\partial \ff t'} \Omega_{\ff t, \ff U} 
[\ff \Sigma_{\ff t' , \ff U}] = 0
\: .
\label{eq:eu}
\end{equation}
The type of the approximation is determined by the choice of the reference system,
i.e.\ by the cluster size $L_{\rm c}$, and by the number of bath degrees of freedom 
$L_{\rm b}$. 
The DMFT is obtained for $L_{\rm c}=1$ and $L_{\rm b}=\infty$, for the C-DMFT
one needs $L_{\rm c}>1$, and the VCA is specified by the choice $L_{\rm c}>1$ 
and $L_{\rm b}=0$ (see Fig.\ \ref{fig:apps}). 
Clearly, there are more possibilities.
Approximations constructed in this way are dynamic and thermodynamically consistent in
general:
Via the self-energy at the stationary point, they provide information on the one-particle 
excitations and an explicit, though approximate, expression for a thermodynamic potential 
from which all static quantities of interest can be derived.

Contrary to the Ritz variational approach, the SFT cannot predict exact upper bounds 
for the grand potential.
From the Ritz principle, or from its generalization for arbitrary temperatures \cite{Mer65},
one has $\Omega_{\ff t, \ff U} [\rho] \ge \Omega_{\ff t, \ff U}$, i.e.\ the grand potential
at an arbitrary density matrix $\rho$ represents an upper bound of the exact grand potential 
$\Omega_{\ff t, \ff U}$.
On the other hand, nothing prevents that
$\Omega_{\ff t, \ff U} [\ff \Sigma_{\ff t' , \ff U}] < \Omega_{\ff t, \ff U}$
for some $\ff t'$ within the SFT.

This raises a number of questions which are addressed in the present paper:
(i)
Is there more than a single stationary point of the self-energy functional, i.e.\
is there more than a single solution of the Euler equation (\ref{eq:eu})?
(ii)
If this is the case, which one is to be preferred?
(iii)
Comparing two different approximations resulting from two different choices of the 
reference system, which one is more reliable? 

These are questions that refer quite generally to any variational principle that 
does not share with the Ritz principle the ``upper-bound property''.
It will be argued that always taking the stationary point with the lowest SFT 
grand potential is a strategy that is generally unacceptable.
A different strategy is suggested instead.

The paper is organized as follows:
The next section briefly reviews the basic concepts of the SFT.
An explicit form of the Euler equation (\ref{eq:eu}) is derived in section 
\ref{sec:euler}. 
Section \ref{sec:theorem} presents an analysis of the Euler equation for the 
case of a reference system composed of two decoupled subsystems. 
This forms the basis for the general discussion on the relative quality of 
different approximations and a systematic way to approach the exact solution 
in section \ref{sec:hier}.
The main conclusions are summarized in section \ref{sec:con}.

\section{Self-energy-functional theory}
\label{sec:sft}

The central idea of the self-energy-functional theory (SFT) is to make use of the 
universality of the Luttinger-Ward functional $\Phi_{\ff U}[\ff G]$ \cite{LW60} or 
of its Legendre transform $F_{\ff U} [\ff \Sigma]$:
For a system with Hamiltonian $H=H_{0} (\ff t) + H_{1} (\ff U)$, where $\ff t$ are 
the one-particle and $\ff U$ the interaction parameters, the functional dependence
$F_{\ff U}[\cdots]$ is independent of $\ff t$ \cite{Pot04}.
The grand potential of the system at temperature $T$ and chemical potential $\mu$ 
can be written as a functional of $\ff \Sigma$: 
\begin{equation}
	\Omega_{\ff t, \ff U} [\ff \Sigma] =  \Tr \ln (\ff G_{0, \ff t}^{-1} -
		\ff \Sigma)^{-1} + F_{\ff U} [\ff \Sigma] \: ,
\label{eq:sf}		
\end{equation}
where $\ff G_{\ff t,0} = (i\omega_n + \mu - \ff t)^{-1}$ is the free Green's function
and $\mbox{Tr} \equiv T \sum_{n} e^{i\omega_n 0^+} \tr$ with the usual trace $\tr$ 
and the Matsubara frequencies $\omega_n = (2n+1)\pi T$ for $n=0,\pm 1,\pm 2,\cdots$.
After Legendre transformation, the basic property of the Luttinger-Ward functional reads 
as $T^{-1} \delta F_{\ff U}[\ff \Sigma]/\delta \ff \Sigma = - \ff G_{\ff U}[\ff \Sigma]$.
This implies that Dyson's equation can be derived by functional differentiation, 
$\delta \Omega_{\ff t, \ff U}[\ff \Sigma] / \delta \ff \Sigma = 0$.
Hence, at the physical self-energy $\ff \Sigma = \ff \Sigma_{\ff t,\ff U}$, the grand 
potential is stationary: $\delta \Omega_{\ff t, \ff U} [\ff \Sigma_{\ff t , \ff U}] = 0$.

Due to the universality of $F_{\ff U}[\ff \Sigma]$, one has
\begin{equation}
	\Omega_{\ff t', \ff U} [\ff \Sigma] = F_{\ff U} [\ff \Sigma] + \Tr \ln
	(\ff G_{\ff t',0}^{-1}-\ff \Sigma)^{-1}
\label{eq:sfp}
\end{equation}
for the self-energy functional of a so-called ``reference system'' which
is given by a Hamiltonian with the same interaction part $\ff U$ but modified
one-particle parameters $\ff t'$: $H'=H_{0} (\ff t') + H_{1} (\ff U)$.
The reference system has different microscopic parameters but is taken to be in the 
same macroscopic state, i.e.\ at the same temperature $T$ and the same chemical 
potential $\mu$.
By a proper choice of its one-particle part, the problem posed by the reference 
system $H'$ can be much simpler than the original problem posed by $H$, 
such that the self-energy of the 
reference system $\ff \Sigma_{\ff t',\ff U}$ can be computed exactly within a
certain subspace of parameters $\ff t'$.
Combining Eqs.\ (\ref{eq:sf}) and (\ref{eq:sfp}), one can eliminate the functional
$F_{\ff U}[\ff \Sigma]$.
Inserting as a trial self-energy the self-energy of the reference system then
yields:
\begin{equation}
	\Omega_{\ff t, \ff U} [\ff \Sigma_{\ff t',\ff U}] 
	= 
	\Omega_{\ff t', \ff U}
	+
	\Tr \ln	(\ff G_{\ff t,0}^{-1}-\ff \Sigma_{\ff t',\ff U})^{-1}
	- 
	\Tr \ln \ff G_{\ff t',\ff U} \: ,
\label{eq:ocalc}	
\end{equation}
where $\Omega_{\ff t', \ff U}$ and $\ff G_{\ff t',\ff U}$ are the grand
potential and the Green's function of the reference system.
This shows that the self-energy functional can be evaluated exactly on the subspace
of trial self-energies that are generated by the reference system.
Solutions of Eq.\ (\ref{eq:eu}) represent stationary points of the functional on
this subspace.
For further details of the approach and for different applications see Refs.\ 
\cite{PAD03,DAH+04,AAPH05,Pot03a,Pot05,Pot04,Pot03b,Poz04,KMOH04,AEvdLP+04,ASE05,SLMT05,AA05,Ton05,IKSK05a,IKSK05b}.

\section{SFT Euler equation}
\label{sec:euler}

Let $\{ | \alpha \rangle \}$ denote the orthonormal set of one-particle basis states.
Then $t_{\alpha \beta}$ are the elements of $\ff t$,
$G_{\ff t,\ff U;\alpha\beta}(i\omega_n) = \langle \langle c_\alpha ; c^\dagger_\beta 
\rangle \rangle_{\ff t,\ff U}$
are the elements of $\ff G_{\ff t,\ff U}=\ff G_{\ff t,\ff U}(i\omega_n)$, etc.
Carrying out the partial differentiation $\partial / \partial \ff t'$
in Eq.\ (\ref{eq:eu}), one arrives at 
\begin{equation}
   T \sum_{n , \alpha\beta}
   \left( 
   \frac{1}{{\ff G}_{\ff t,0}^{-1}(i\omega_n) - {\ff \Sigma}_{\ff t',\ff U}(i\omega_n)} 
   -  {\ff G}_{\ff t',\ff U}(i\omega_n) 
   \right)_{\beta \alpha} 
   \frac{\partial \Sigma_{\ff t',\ff U;\alpha\beta}(i\omega_n)}
        {\partial {{\ff t}'}}
   = 0 
   \: .
\label{eq:euler}
\end{equation}
Note that there are as much (non-linear) equations as there are unknowns 
$t'_{\alpha \beta}$.
The Euler equation (\ref{eq:euler}) can be derived from the representation 
(\ref{eq:ocalc}) for the SFT grand potential by using the (exact) relation
$\partial \Omega_{\ff t',\ff U} / \partial t'_{\alpha\beta} =
\langle c_\alpha^\dagger c_\beta \rangle_{\ff t',\ff U} = T \sum_n e^{i\omega_n0^+}
(i\omega_n + \mu - \ff t' - {\ff \Sigma}_{\ff t',\ff U})^{-1}_{\beta \alpha}$.
The Euler equation is trivially fulfilled for 
${\ff t}' = \ff t$ since 
$({\ff G}_{\ff t,0}^{-1} - {\ff \Sigma}_{\ff t,\ff U})^{-1} = {\ff G}_{\ff t,\ff U}$.
In all practical situations, however, the point ${\ff t}' = \ff t$ does not belong
to the parameter space of the reference system since the $\ff t'$ must be chosen
such that the problem posed by $H'$ is exactly solvable.

The term $\partial \Sigma_{\ff t',\ff U;\alpha\beta}(i\omega_n) / \partial {\ff t}'$ 
may be considered as a projector.
In the space of self-energies,
$\partial \Sigma_{\ff t',\ff U;\alpha\beta}(i\omega_n) / \partial {\ff t}'$ 
is a vector tangential to the hypersurface of ${\bf t}'$ representable trial 
self-energies ${\ff \Sigma}_{\bf t',\ff U}$. 
Hence, the Euler equation Eq.\ (\ref{eq:euler}) determines the self-energy
from its exact conditional equation (Dyson's equation) but projected onto that
hypersurface by taking the scalar product with the projectors 
$\partial {\ff \Sigma}_{\ff t',\ff U} / \partial {\ff t}'$.

The projectors can be determined more explicitly by carrying out the $\ff t'$
differentiation.
Writing $\ff \Sigma = {\ff \Sigma}_{\ff t',\ff U}$ and 
$\ff G' = {\ff G}_{\ff t',\ff U}$ for short, one has 
\begin{equation}
   \Sigma_{\alpha\beta}(i\omega_n) =
   (i\omega_n + \mu) \delta_{\alpha\beta} - t'_{\alpha\beta} - 
   {G'}_{\alpha\beta}^{-1}(i\omega_n) \: 
\label{eq:dyhprime}
\end{equation}
from Dyson's equation for the reference system.
Hence:
\begin{equation}
   \frac{\partial \Sigma_{\alpha\beta}(i\omega_n)}
        {\partial {t'_{\rho\sigma}}} 
   =
   - \delta_{\alpha\rho} \delta_{\beta\sigma}
   - \frac{\partial {G'}_{\alpha\beta}^{-1}(i\omega_n)}{\partial t'_{\rho\sigma}}   
   \: .
\end{equation}
Making use of the relation $\delta ({\bf B} - {\bf A})_{ij}^{-1} / \delta A_{mn} 
= ({\bf B} - {\bf A})_{im}^{-1} ({\bf B} - {\bf A})_{nj}^{-1}$ which holds for two
not necessarily commuting matrices $\bf A$ and ${\bf B}$, 
\begin{equation}
   \frac{\partial \Sigma_{\alpha\beta}(i\omega_n)}
        {\partial {t'_{\rho\sigma}}} 
    = - \delta_{\alpha\rho} \delta_{\beta\sigma}
   + \sum_{\mu\nu}{G'}_{\alpha\mu}^{-1}(i\omega_n)
   \frac{\partial G'_{\mu\nu}(i\omega_n)}{\partial t'_{\rho\sigma}}   
  {G'}_{\nu\beta}^{-1}(i\omega_n) \: .
\end{equation}

To calculate the linear response of the Green function when varying the one-particle
parameters, the $S$ matrix shall be introduced by the definition
\begin{equation}
   S_{\ff t',\ff U}(1/T) 
   = 
   \exp(H_1(\ff U) / T) \exp( -( H_0(\ff t') + H_1(\ff U) - \mu N)/T) \: .
\end{equation} 
$N$ is the particle number operator.
Note that as compared to the conventional definition for $S$, the roles of the free and 
of the interaction part of the Hamiltonian are interchanged, i.e.\ $H_0(\ff t') - \mu N$ 
is considered as a perturbation here.
With the time ordering operator ${\cal T}$, the $S$ matrix can be written as
\begin{equation}
  S_{\ff t',\ff U}(1/T) 
  = 
  {\cal T} \exp \left( 
  - \int_0^{1/T} d \tau \sum_{\rho\sigma} (t'_{\rho\sigma} - \mu \delta_{\rho\sigma})
  \widetilde{c}_\rho^\dagger(\tau) \widetilde{c}_\sigma(\tau)
  \right) \: .
\end{equation}
Here the notation $\widetilde{O}(\tau)=\exp(H_1(\ff U)\tau) O \exp(-H_1(\ff U)\tau)$
is used: The imaginary time dependence is due to $H_1(\ff U)$ only.

Now the Green's function
$G'_{\mu\nu}(i\omega_n) = \int_0^{1/T} d\tau \: e^{i\omega_n\tau}G'_{\mu\nu}(\tau)$
can be written as 
\begin{equation}
  G'_{\mu\nu}(\tau) = - \frac{
  {\rm tr} \: e^{-H_1(\ff U)/T} {\cal T} \: S \: \widetilde{c}_\mu(\tau) 
  \widetilde{c}_\nu^\dagger(0)
  }{
  {\rm tr} \: e^{-H_1(\ff U)/T} \: S 
  } 
\label{eq:grep}  
\end{equation}
with $S = S_{\ff t',\ff U}(1/T)$ for short.
Again, the (interchanged) interaction representation is used with the $\tau$ 
dependence of $\widetilde{c}_\mu(\tau)$ being 
due to $H_1(\ff U)$. 
The ${\ff t}'$ dependence of the Green's function is due to $S$ only:
\begin{equation}
  \frac{\partial S}{\partial t'_{\rho\sigma}} =
  - {\cal T} \int_0^{1/T} d \tau \: 
  \widetilde{c}_\rho^\dagger(\tau) \widetilde{c}_\sigma(\tau) \: S
\end{equation}
Using this in Eq.\ (\ref{eq:grep}),
\begin{equation}
  \frac{\partial G'_{\mu\nu}(\tau)}{\partial t'_{\rho\sigma}} = -
  \int_0^{1/T} d \tau' \:  L'_{\mu \sigma \rho \nu}(\tau,\tau';\tau'_+,0)
\end{equation}
where $L'$ is a two-particle dynamical correlation function of the reference system $H'$:
\begin{eqnarray}
  L'_{\rho_1\rho_2\rho_3\rho_4}(\tau_1,\tau_2;\tau_3,\tau_4) &=& 
  \langle {\cal T}
  c_{\rho_1}(\tau_1) c_{\rho_2}(\tau_2) 
  c_{\rho_3}^\dagger(\tau_3) c_{\rho_4}^\dagger(\tau_4) \rangle 
\nonumber \\ && \mbox{} \hspace{-20mm}
    - \langle {\cal T} c_{\rho_1}(\tau_1) c_{\rho_4}^\dagger(\tau_4) 
      \rangle \:
      \langle {\cal T} c_{\rho_2}(\tau_2) c_{\rho_3}^\dagger(\tau_3) 
      \rangle 
\label{eq:ldef}
\end{eqnarray}
and $\tau_+=\tau+0^+$.
Here the average and the time dependence is due to the {\em full} Hamiltonian: 
$O(\tau)=\exp((H'-\mu N)\tau) O \exp(-(H'-\mu N)\tau)$.
Defining
\begin{equation}
  L'_{\mu \sigma \rho \nu}(i\omega_n) =
  \int_0^{1/T} \int_0^{1/T} d \tau d \tau' \: e^{i\omega_n \tau} \:
  L'_{\mu \sigma \rho \nu}(\tau,\tau';\tau'_+,0) \; ,
\end{equation}
one has
\begin{equation}
  \frac{\partial G'_{\mu\nu}(i\omega_n)}{\partial t'_{\rho\sigma}} = -
  L'_{\mu \sigma \rho \nu}(i\omega_n)
\end{equation}
and thus
\begin{equation}
   \frac{\partial \Sigma_{\alpha\beta}(\omega)}
        {\partial {t'_{\rho\sigma}}} 
   = - \delta_{\alpha\rho} \delta_{\beta\sigma}
   - \sum_{\mu\nu}{G'}_{\alpha\mu}^{-1}(\omega)
   L'_{\mu \sigma \rho \nu}(\omega)
  {G'}_{\nu\beta}^{-1}(\omega) \: .
\label{eq:proj}
\end{equation}

\begin{figure}[t]
\centerline{\includegraphics[width=0.8\textwidth]{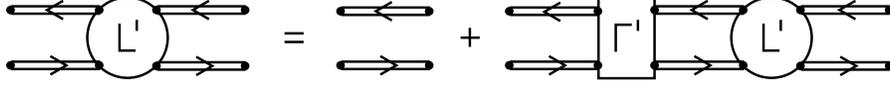}}
\caption{
Diagrammatic representation of the relation between the two-particle Green's
function $L'$ and the two-particle self-energy $\Gamma'$.
Double lines represent the one-particle Green's function $\ff G'$.
}
\label{fig:2sig}
\end{figure}

\begin{figure}[t]
\centerline{\includegraphics[width=0.65\textwidth]{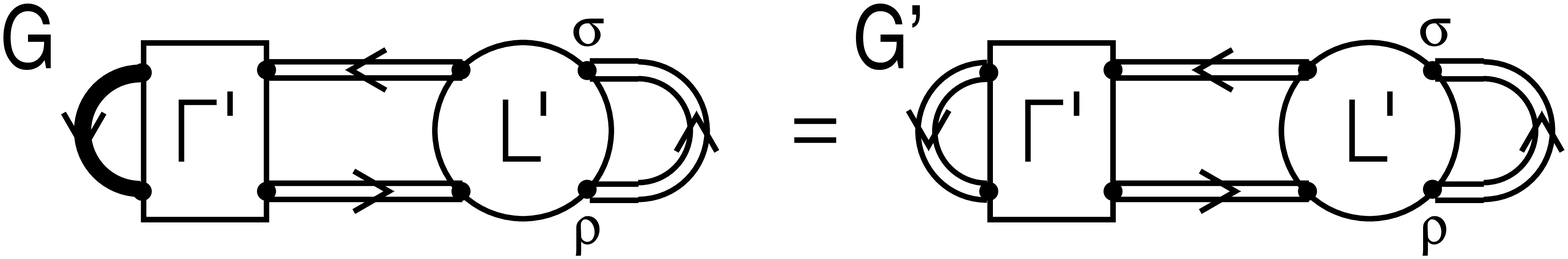}}
\caption{
Diagrammatic representation of the Euler equation (\ref{eq:finaleuler}).
All variables except for $\rho$ and $\sigma$ are summed over.
Rectangular box: two-particle self-energy $\Gamma'$.
Circle and adjacent double lines: two-particle response function $L'$.
Double line: Green's function of the reference system ${\ff G}'$.
The solid line denotes
the approximate Green's function $({\ff G}_{\ff t,0}^{-1} - {\ff \Sigma})^{-1}$
of the original system.
}
\label{fig:euler}
\end{figure}

Introducing the two-particle self-energy of $H'$ (see Fig.\ \ref{fig:2sig}),
\begin{equation}
  \Gamma'_{\rho_1\rho_2\rho_3\rho_4}(\omega_1,\omega_2) = \frac{1}{T} \:
  \frac{\delta \Sigma_{\rho_1\rho_4}(\omega_1)[{\ff G}']}
       {\delta G'_{\rho_3\rho_2}(\omega_2)} \: ,
\end{equation}
where the functional of ${\ff G}'$ is the functional derivative of the 
Luttinger-Ward functional,
this can also be written as
\begin{equation}
  \frac{\partial \Sigma_{\alpha\beta}(i\omega_n)}{\partial {t'_{\rho\sigma}}} 
  = - T \sum_{m} \sum_{\mu\nu} 
  \Gamma'_{\alpha\nu\mu\beta}(i\omega_n,i\omega_m)
  L'_{\mu \sigma \rho \nu}(i\omega_m) \: ,
\end{equation}
which finally yields the Euler equation in the form (see Fig.\ \ref{fig:euler}):
\begin{equation}
   0 
   =
   T^2 \sum_{nm} \sum_{\alpha\beta\mu\nu}
   \left[ 
   \frac{1}{{\ff G}_{\ff t, 0}^{-1}(i\omega_n) - {\ff \Sigma}(i\omega_n)}
   - 
   {\ff G}'(i\omega_n) 
   \right]_{\beta \alpha}
   \Gamma'_{\alpha\nu\mu\beta}(i\omega_n,i\omega_m)
   L'_{\mu \sigma \rho \nu}(i\omega_m) \: .
\label{eq:finaleuler}
\end{equation}

If the system $H'$ not only consists of one-particle orbitals belonging to $H$ but also 
includes additional (uncorrelated) bath orbitals, one has to be careful with the orbital
indices. Throughout the above derivation, $\alpha,\beta,\gamma,\delta$ refer to the orbitals 
of $H$, while $\mu,\nu,\rho,\sigma$ refer to the orbitals of $H'$. 
An orbital index $\mu$ of $H'$ runs over $\mu=\alpha=1,2,...,M$ 
(the correlated orbitals of $H'$ which are identified with corresponding orbitals of $H$, 
where usually $M \to \infty$) and additionally over $\mu=M+1,M+2,...$ 
(the uncorrelated bath orbitals).
${\ff G}'$ denotes the Green function of the reference system with the elements 
$G'_{\mu \nu}(i\omega_n)$.
On the correlated orbitals $\mu=\alpha$, $\nu=\beta$ one has
$G'_{\alpha\beta}(i\omega_n) = {G}_{\ff U;\alpha\beta}(i\omega_n) [{\ff \Sigma}]$.
Recall that ${\ff G}_{\ff U}[{\ff \Sigma}]$ is the inverse functional of 
${\ff \Sigma}_{\ff U}[{\ff G}]$
which only includes the propagators between correlated sites $\alpha$ and $\beta$.
When additional uncorrelated sites are considered, the equation (\ref{eq:dyhprime}) 
is not the complete Dyson equation in $H'$ but only the block with $\alpha$, $\beta$ 
elements. Note that 
${\ff G'}^{-1}$ means matrix inversion with respect to all orbitals of $H'$.

As compared with the DMFT or the C-DMFT self-consistency equation,
the SFT Euler equation (\ref{eq:finaleuler}) is more complicated as it involves 
dynamical {\em two-particle} correlation functions of the reference system.
As the reference system is assumed to be exactly solvable these are accessible,
in principle.
For practical calculations, a modified version of the Euler equation has been
suggested \cite{Poz04} and shown to allow for an extremely precise determination 
of a stationary point of the self-energy functional.
For the purpose of a general discussion, however, the form (\ref{eq:finaleuler}) is
more useful.

\section{A theorem on decoupled reference systems}
\label{sec:theorem}

Let a reference system consist of two subsystems $A$ and $B$.
Subsystem $A$ is defined as the set of orbitals $| \alpha_A \rangle$, 
and subsystem $B$ is given by the rest of the orbitals $| \alpha_B \rangle$, 
i.e.\ the complete (and orthonormal) one-particle basis is
$\{ | \alpha_A \rangle , | \alpha_B \rangle \}$.
Typically, $A$ and $B$ are given by two disjoint sets of sites.
The Hamiltonian of the reference system can be written as
\begin{equation}
  H' = H'_A + H'_B + H'_{AB} \; ,
\end{equation}
where $H'_A$ only acts in the Fock space of $\{ | \alpha_A \rangle \}$ and
$H'_B$ in the Fock space of $\{ | \alpha_B \rangle \}$. Hence, the commutator
$[H'_A,H'_B]_-=0$.
$H'_{AB}$ is a term which couples the dynamics of the two subsystems and is 
assumed to be a one-particle operator, i.e.\ a coupling due to a two-particle 
interaction part of the Hamiltonian is excluded. 
This is satisfied, for example, in case of the Hubbard model if the subsystems
are given on disjoint sets of sites as the Hubbard interaction is local.
The coupling term can then be written as $H'_{AB}=H_{0}(\ff V)$ where 
$\ff V$ is the matrix of one-particle coupling parameters 
$V_{\alpha_A \alpha_B}$ with $\alpha_A \in A$ and $\alpha_B \in B$.

A given reference system specifies a certain space of trial self-energies
for the self-energy functional and thereby a certain approximation.
What is the relation between an 
approximation given by the reference system $H' = H'_A + H'_B + H'_{AB}$
and an approximation given by the decoupled system $H'' = H'_A + H'_B$?
With the above preconditions, the following theorem holds:
Any stationary point of the self-energy functional on the subspace of self-energies
defined by the decoupled system $H''$ is also a stationary point on the subspace of 
self-energies defined by the coupled system $H'$, namely at $\ff V = 0$.
Writing $H' = H_0(\ff t') + H_1(\ff U) = H_0(\ff t'') + H_0(\ff V) + H_1(\ff U)$
and $H''=H_0(\ff t'') + H_1(\ff U)$ and 
\begin{equation}
\ff t' = \left( 
\begin{array}{cc}
\ff t'_{AA} & \ff 0 \\
\ff 0 & \ff t'_{BB}
\end{array}
\right)
+ \left( 
\begin{array}{cc}
\ff 0 & \ff t'_{AB} \\
\ff t'_{BA} & \ff 0
\end{array}
\right)
= \ff t'' + \ff V \; ,
\end{equation}
the theorem is:
\begin{equation}
\frac{\partial}{\partial \ff t''} \Omega_{\ff t, \ff U} [\ff \Sigma_{\ff t'' , \ff U}] = 0
\qquad
\Rightarrow
\qquad
\frac{\partial}{\partial \ff t'} \Omega_{\ff t, \ff U} [\ff \Sigma_{\ff t' , \ff U}] 
\bigg|_{\ff V=0}= 0 \: .
\end{equation}
While the theorem is not trivial, it complies with intuitive expectations:
Going from a more simple reference system to a more complicated reference system 
with more degrees of freedom coupled, should generate a new stationary
point with $\ff V \ne 0$; the ``old'' stationary point with $\ff V = 0$, however, 
is still a stationary point in the ``new'' reference system.
Therefore, coupling more and more degrees of freedom, 
introduces more and more stationary
points of the self-energy functional, and none of the old ones is ``lost''.

For the proof the results of the preceeding section are needed, in particular the 
representation of the projector 
$\partial \Sigma_{\alpha\beta}(i\omega_n) / \partial {t'_{\rho\sigma}}$ 
in Eq.\ (\ref{eq:proj}).
One has to distinguish between the following different cases:

(i) $\alpha,\beta \in A$ and $\rho,\sigma \in B$:
For $\ff V = 0$ the Green's function as well as its inverse does not couple
orbitals of different subsystems, e.g.\ ${G'}_{\alpha\mu}^{-1}(i\omega_n) = 0$ 
if $\alpha \in A$ and $\mu \in B$.
Hence, in Eq.\ (\ref{eq:proj}) there can be non-zero contributions 
for $\mu,\nu \in A$ only.
Since $\rho,\sigma \in B$ and $\ff V = 0$, the first term of the two-particle
Green's function in Eq.\ (\ref{eq:ldef}) decouples and thus:
\begin{eqnarray}
  L'_{\mu \sigma \rho \nu}(\tau,\tau';\tau'_+,0) 
&=&
  \langle {\cal T}
  c_{\mu}(\tau) c_{\sigma}(\tau') 
  c_{\rho}^\dagger(\tau'_+) c_{\nu}^\dagger(0) 
  \rangle 
\nonumber \\ 
&-& 
  \langle {\cal T} c_{\mu}(\tau) c_{\nu}^\dagger(0) \rangle \:
  \langle {\cal T} c_{\sigma}(\tau') c_{\rho}^\dagger(\tau'_+) \rangle 
\nonumber \\ 
&=&
  \langle {\cal T} c_{\mu}(\tau) c_{\nu}^\dagger(0) \rangle \:
  \langle {\cal T} c_{\sigma}(\tau') c_{\rho}^\dagger(\tau'_+) \rangle 
\nonumber \\ 
&-&
  \langle {\cal T} c_{\mu}(\tau) c_{\nu}^\dagger(0) \rangle \:
  \langle {\cal T} c_{\sigma}(\tau') c_{\rho}^\dagger(\tau'_+) \rangle 
  = 0
  \: ,
\end{eqnarray}
which implies 
$\partial \Sigma_{\alpha\beta}(i\omega_n) / \partial {t'_{\rho\sigma}}=0$.
The case $\alpha,\beta \in B$ and $\rho,\sigma \in A$ can be treated
analogously.

(ii) $\alpha,\beta,\rho \in A$ and $\sigma \in B$:
In Eq.\ (\ref{eq:proj})
there can be non-zero contributions for $\mu,\nu \in A$ only.
For $\ff V=0$ the Green's function $L'$ decouples and vanishes
since $\langle c_{\sigma}(\tau') \rangle = 0$ for fermions.
Consequently, 
$\partial \Sigma_{\alpha\beta}(i\omega_n) / \partial {t'_{\rho\sigma}}=0$.
The same type of reasoning applies to the cases
$\alpha,\beta,\sigma \in A$, $\rho \in B$
and
$\alpha,\sigma,\rho \in A$, $\beta \in B$
and
$\beta,\sigma,\rho \in A$, $\alpha \in B$ as well as to
those cases with the roles of $A$ and $B$ interchanged.

(iii) $\alpha,\rho \in A$ and $\beta, \sigma \in B$:
In this case, $\mu \in A$ and $\nu \in B$ which implies (for $\ff V=0$)
that the second term of the two-particle Green's function in Eq.\ (\ref{eq:ldef}) 
vanishes and the first term decouples:
\begin{eqnarray}
  L'_{\mu \sigma \rho \nu}(\tau,\tau';\tau'_+,0) 
&=&
  \langle {\cal T}
  c_{\mu}(\tau) c_{\sigma}(\tau') 
  c_{\rho}^\dagger(\tau'_+) c_{\nu}^\dagger(0) 
  \rangle 
\nonumber \\ 
&=& 
  -
  \langle {\cal T} c_{\mu}(\tau) c_{\rho}^\dagger(\tau'_+) \rangle \:
  \langle {\cal T} c_{\sigma}(\tau') c_{\nu}^\dagger(0) \rangle 
\nonumber \\ 
&=& 
  -
  G'_{\mu\rho}(\tau-\tau')
  G'_{\sigma\nu}(\tau')
  \; .
\end{eqnarray}
This yields:
\begin{eqnarray}
  L'_{\mu \sigma \rho \nu}(i\omega_n) 
  &=&
  -
  \int_0^{1/T} \int_0^{1/T} d \tau d \tau' \: e^{i\omega_n \tau} \:
  G'_{\mu\rho}(\tau-\tau')
  G'_{\sigma\nu}(\tau')
\nonumber \\
  &=&
  -
  G'_{\mu\rho}(i\omega_n)
  G'_{\sigma\nu}(i\omega_n)
\end{eqnarray}
and thus
\begin{equation}
   \frac{\partial \Sigma_{\alpha\beta}(i\omega_n)}
        {\partial {t'_{\rho\sigma}}} 
   = - \delta_{\alpha \rho} \delta_{\beta \sigma}
   + \sum_{\mu\nu}{G'}_{\alpha\mu}^{-1}(i\omega_n)
   G'_{\mu\rho}(i\omega_n)
  G'_{\sigma\nu}(i\omega_n) 
  {G'}_{\nu\beta}^{-1}(i\omega_n) = 0\: .
\end{equation}
Analogously, the projector vanishes if 
$\alpha,\rho \in B$ and $\beta, \sigma \in A$.

(iv) In the case $\alpha,\sigma \in A$ and $\beta, \rho \in B$ and, 
analogously, for $\alpha,\sigma \in B$ and $\beta, \rho \in A$, one is led 
to anomalous correlation functions of the form $\langle c \, c \rangle$ which 
vanish if spontaneous U(1) symmetry breaking is excluded as it is done in 
the derivation of the Euler equation (\ref{eq:eu}) from the very beginning.
As a consequence, one has 
$\partial \Sigma_{\alpha\beta}(i\omega_n) / \partial {t'_{\rho\sigma}}=0$ 
in this case, too.

Thereby all possibilities have been enumerated with the exception of
the two cases $\alpha,\beta,\rho,\sigma \in A$ and 
$\alpha,\beta,\rho,\sigma \in B$.
Here, there is no reason for the projector  
$\partial \Sigma_{\alpha\beta}(i\omega_n) / \partial {t'_{\rho\sigma}}$ 
to vanish even if $\ff V=0$.

These last two cases in fact correspond to variations on the space of 
trial self-energies given by the decoupled system $H''$. 
If there is a stationary point $\ff \Sigma(\ff t'')$ on this smaller space, 
this must necessarily represent a stationary point on the larger space given 
by $H'$, too:
Namely, {\em in the additional cases to be considered}, the Euler equation is 
fulfilled trivially since, as shown above, the projector 
$\partial \Sigma_{\alpha\beta}(i\omega_n) / \partial {t'_{\rho\sigma}}$
vanishes.
Summing up, this shows that any stationary point of the self-energy functional 
on the smaller subspace is also a stationary point on the larger subspace of 
the coupled reference system, namely at $\ff V = 0$.
This proves the theorem.

Put in another way, the theorem states that as a function of a parameter 
(set of parameters) $\ff V$ coupling two separate subsystems, 
\begin{equation}
\Omega_{\ff t, \ff U} [\ff \Sigma_{\ff t''+\ff V, \ff U}] 
=
\Omega_{\ff t, \ff U} [\ff \Sigma_{\ff t''+0, \ff U}] 
+
{\cal O}(\ff V^2) \: , 
\label{eq:exp}
\end{equation}
provided that the functional is stationary at $\ff \Sigma_{\ff t'', \ff U}$ 
{\em when varying $\ff t''$ only} (this restriction makes the theorem non-trivial). 

\section{Hierarchy of stationary points}
\label{sec:hier}

\begin{figure}[t]
\centerline{\includegraphics[width=0.77\textwidth]{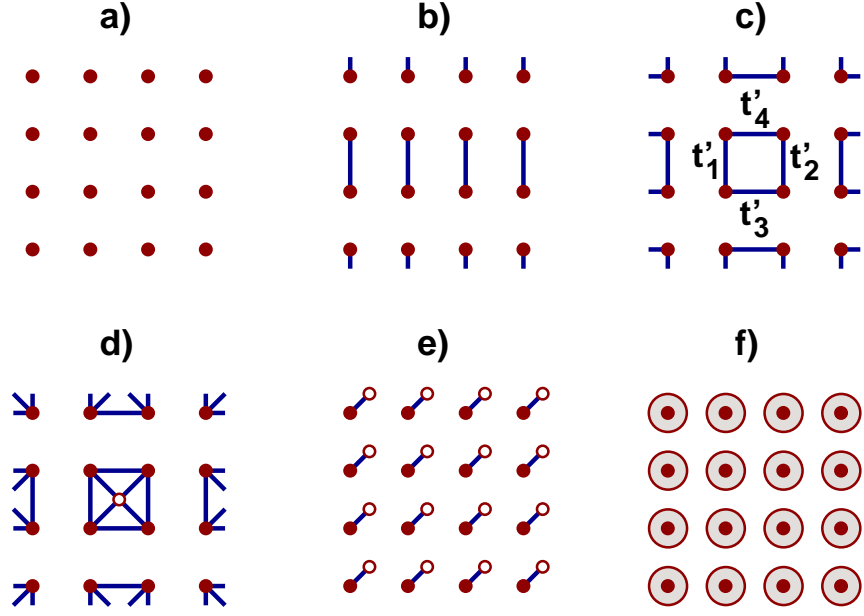}}
\caption{
Different possible reference systems with the same interaction as the
single-band Hubbard model on a square lattice.
Filled circles: correlated sites with $U$ as in the Hubbard model.
Open circles: uncorrelated ``bath'' sites with $U=0$.
Lines: nearest-neighbor hopping.
Big circles: continuous bath consisting of $L_{\rm b} = \infty$ bath sites.
Reference systems $H'_{\rm a}$, $H'_{\rm b}$, $H'_{\rm c}$ generate variational cluster approximations (VCA),
$H'_{\rm e}$ a dynamical impurity approximation, $H'_{\rm f}$ the DMFT, and
$H'_{\rm d}$ an intermediate approximation (VCA with one additional bath site
per cluster).
}
\label{fig:ref}
\end{figure}

To be explicit and to simplify the discussion, the single-band Hubbard model 
\begin{equation}
  H = - t \sum_{\langle ij \rangle,\sigma} c_{i\sigma}^\dagger c_{j\sigma}
  + U \sum_i n_{i\uparrow} n_{i\downarrow}
\end{equation}
on a two-dimensional square lattice with nearest-neighbor hopping is considered 
in the following. 
Nevertheless, the discussion is completely general and applies to arbitrary 
correlated fermionic lattice models with local interactions.
A possible reference system $H'$ must have the same local Hubbard interaction;
the hopping part, however, can be modified arbitrarily.
A number of different reference systems are shown in Fig.\ \ref{fig:ref}.

To discuss a first consequence of the theorem, one should distinguish between
``trivial'' and ``non-trivial'' stationary points for a given reference system. 
A stationary point is referred to as ``trivial'' if the one-particle parameters
are such that the reference system decouples into smaller subsystems.
If, at a stationary point, all degrees of freedom (sites) are still coupled to 
each other, the stationary point is called ``non-trivial''.
In fact, all numerical results that have been obtained so far
\cite{PAD03,DAH+04,AAPH05,Pot03a,Pot05,Pot04,Pot03b,Poz04,KMOH04,AEvdLP+04,ASE05,SLMT05,AA05,Ton05,IKSK05a,IKSK05b}
show that there is at least a single non-trivial stationary point for any reference 
system.

Once this is assumed to be true, then, for a given reference system, there 
must be {\em several} stationary points.
Consider the reference system $H'_{\rm c}$ in Fig.\ \ref{fig:ref}, for example.
Here, there are four intra-cluster nearest-neighbor hopping parameters which 
are treated as independent variational parameters.
A non-trivial stationary point would be a stationary point with 
$t'_1, t'_2, t'_3, t'_4 \ne 0$ (or $t'_1, t'_2, t'_3 \ne 0$, $t'_4=0$).
A second stationary point is then found for $t'_3=t'_4=0$ and some $t'_1, t'_2 \ne 0$
since, according to the theorem, this corresponds to a non-trivial stationary
point generated by the reference system $H'_{\rm b}$.
Another stationary point is obtained with $t'_1 = t'_2 = t'_3 = t'_4 = 0$
since this corresponds to a stationary point generated by $H'_{\rm a}$.
(Note that the one-particle energies $\varepsilon'_i \equiv t'_{ii}$ are variational 
parameters, too.)
This shows that within a given approximation, i.e.\ for a given reference system,
a non-trivial stationary point has always to be compared with several (on that 
level) trivial stationary points.

Now, it is important to note that a stationary point of the self-energy functional 
$\Omega_{\ff t,\ff U}[\ff \Sigma_{\ff t',\ff U}]$ is not necessarily a minimum.
In general, a saddle point is found.
This is demonstrated, for example, by the calculation in Ref.\ \cite{Pot03a}.
Furthermore, there is no reason why, for a given reference system, the SFT
grand potential at a non-trivial stationary point should be lower than the 
SFT grand potential at a trivial one.
And finally, it cannot be ensured that the SFT grand potential, evaluated at
a given trial self-energy, is always higher than the exact grand potential,
i.e.\ $\Omega_{\ff t,\ff U}[\ff \Sigma_{\ff t',\ff U}] < \Omega_{\ff t,\ff U}$ 
may be possible.
This stands in sharp contrast to the Ritz variational principle.
The fact that the spectrum of the Hamiltonian (after a constant energy shift) 
is always positive definite guarantees the upper-bound property
$\langle \Psi | H | \Psi \rangle \ge E_0$.
It is not surprising that this upper-bound property is lost within the SFT
as the approach does not refer to wave functions $|\Psi \rangle$ at all.
This is probably characteristic for {\rm any} dynamical variational approach,
i.e.\ for variational approaches based on time-dependent correlation functions, 
Green's functions, self-energies, etc.

Consider the case where there is a non-trivial stationary point and a number
of trivial stationary points for a given reference system.
Despite the above reasoning, an intuitive strategy to decide between two 
stationary points is to always take the one with the lower grand potential 
$\Omega_{\ff t,\ff U}[\ff \Sigma_{\ff t',\ff U}]$.
A sequence of reference systems (e.g. $H'_{\rm a}$, $H'_{\rm b}$, $H'_{\rm c}$,
...) in which more and more degrees of freedom are coupled and which eventually 
recovers the original system $H$ itself, shall be called a ``systematic'' sequence 
of reference systems.
For such a systematic sequence, the suggested strategy will produce a series of 
stationary points with monotonously decreasing grand potential.
This is reminiscent of the Ritz principle. 
Furthermore, by comparing the trends of the SFT grand potential for two 
stationary points as functions of an external parameter, one can easily 
identify level crossings as well as continuous or discontinuous phase transitions
and interprete them consistently within the framework of equilibrium thermodynamics.

Unfortunately, however, the strategy is useless because it cannot ensure that a 
systematic sequence of reference systems generates a systematic sequence of 
approximations as well:
Within the SFT, one cannot ensure that the respective {\em lowest} grand potential 
in a systematic sequence of reference systems and corresponding stationary 
points converges to the exact grand potential.
This means that despite the fact that the complexity of the reference systems
increases, the stationary point with the lowest SFT grand potential could be
a trivial stationary point, i.e.\ could be associated with a very simple reference
system only (like $H'_{\rm a}$ or $H'_{\rm b}$, for example).
Such an approximation must be considered as poor since the exact conditional 
equation for the self-energy is projected onto a very low-dimensional space only.

Therefore, one has to construct a different strategy which necessarily approaches 
the exact solution when following up a systematic sequence of reference systems.
Clearly, this can only be achieved if the following rule is obeyed:
\begin{itemize}
\item
A non-trivial stationary point is always preferred as compared to a trivial one (R0).
\end{itemize}
A non-trivial stationary point at a certain level of approximation, i.e.\ for a
given reference system becomes a trivial stationary point on the next level, i.e.\
in the context of a ``new'' reference system that couples at least two different
units of the ``old'' reference system.
Hence, by construction, the rule R0 implies that the exact result is approached
for a systematic series of reference systems.

\begin{figure}[t]
\centerline{\includegraphics[width=0.95\textwidth]{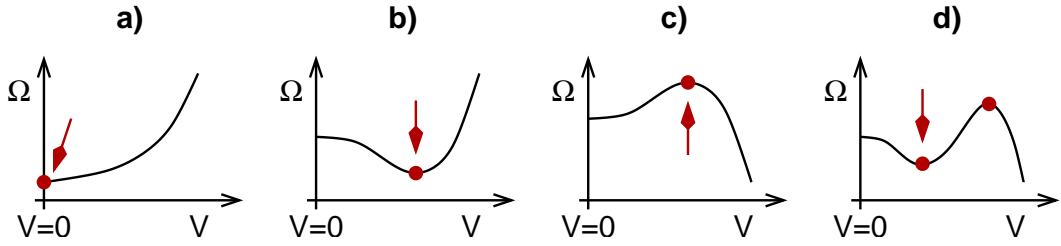}}
\caption{
Schematic trends of the SFT grand potential $\Omega$ as a function of a
variational parameter $V$ which is assumed to couple two different subsystems 
of the reference system, i.e.\
$V=0$ corresponds to the decoupled case and (according to the theorem of
Sec.\ \ref{sec:theorem}) must always represent a (``trivial'') stationary point.
Circles show the stationary points to be considered.
According to R1, the trivial stationary point $V=0$ has to be disregarded in all cases
except for a).
The arrow marks the respective optimum stationary point.
Note that in case c) the SFT grand potential is higher for the non-trivial
as compared to the trivial stationary point.
In case d) the rule R2 applies.
}
\label{fig:om}
\end{figure}

Following the rule (R0), however, may lead to inconsistent thermodynamic 
interpretations for the case that a trivial stationary point has a lower 
grand potential as the non-trivial one.
To avoid this, another rule is necessary:
\begin{itemize}
\item
Trivial stationary points have to be disregarded completely unless there is no
non-trivial one (R1).
\end{itemize}
This automatically ensures that there is at least one stationary point for any
reference system, i.e.\ there is at least one solution at any level of the 
approximation.
Clearly, R1 makes R0 superfluous.

To maintain a thermodynamically consistent picture in case that there are more 
than a single {\em non-trivial} stationary points, one needs the following rule:
\begin{itemize}
\item
Among two non-trivial stationary points for the same reference system, 
the one with lower grand potential has to be preferred (R2).
\end{itemize}

The rules are illustrated by Fig.\ \ref{fig:om} which gives different examples.
Note that the grand potential away from a stationary point does not have a direct
physical interpretation.
Hence, there is no reason to interprete the solution corresponding to 
the maximum in Fig.\ \ref{fig:om}, c) as ``locally unstable''.
The results of Ref.\ \cite{Pot03b} (see Figs.\ 2 and 4 therein) nicely demonstrate
that with the suggested strategy (R1, R2) one can consistently describe 
continuous as well as discontinuous phase transitions.

It should be stressed that the above rules R1 and R2 are unambiguously prescribed 
by the general demands for the possibility of systematic improvement and for 
thermodynamic consistency.
There is no acceptable alternative to this strategy.
Note that the strategy reduces to the standard strategy (always taking the 
solution with lowest grand potential or, for $T=0$, the lowest ground-state 
energy) in case of the Ritz variational principle because here a non-trivial 
stationary point does always have a lower grand potential as compared to a 
trivial one.

There are also some consequences of the strategy which might be considered as 
disadvantageous but must be tolerated:
(i) For a sequence of stationary points that are determined by R1 and R2 from 
a systematic sequence of reference systems, the corresponding sequence of SFT
grand potentials necessarily converges to the exact grand potential but not
necessarily in a monotonous way. 
For example, the exact grand potential might be approached from below or in an 
oscillatory way.
(ii) Given two different approximations specified by two different reference 
systems, it is not possible to decide which one should be regarded as superior
unless both reference systems belong to the same systematic sequence of reference 
systems.
In Fig.\ \ref{fig:ref}, for example, one has $H'_{\rm a} < H'_{\rm b} < H'_{\rm c}
< H'_{\rm d}$ where ``$<$'' stands for ``is inferior compared to''.
Furthermore, $H'_{\rm e} < H'_{\rm f}$ and $H'_{\rm a} < H'_{\rm e}$ but there is 
no relation between $H'_{\rm b}$ and $H'_{\rm e}$, for example.

\section{Conclusions}
\label{sec:con}

A dynamical variational principle is a principle of the form 
$\delta \Omega[\ff X] = 0$ where $\Omega$ is a thermodynamic potential and
$\ff X$ is a dynamical quantity that refers to excitations of the system out
of equilibrium but in the linear-response regime.
A common characteristic of the different dynamical variational principles used 
in solid-state theory \cite{Pot05} is that stationary points are saddle points
rather than minima in general and that the thermodynamic potential at a stationary 
point cannot serve as an upper bound of the true potential.
One of the most famous approximations that can be constructed in this context
is the dynamical mean-field theory and, in fact, there is no general proof 
(for finite-dimensional lattice models) that 
$\Omega_{\rm DMFT} \ge \Omega_{\rm exact}$ so far.

Having these problems in mind, it becomes questionable how to judge on the 
{\em relative} quality of two different approximations resulting from two 
different stationary points of a dynamical variational principle.
It has been shown here that at least within the context of the 
self-energy-functional theory there is an answer to this question which is 
prescribed by demanding approximations to be thermodynamically consistent
as well as systematic and which is summarized by the rules R1 and R2 in 
Sec.\ \ref{sec:hier}.
It has turned out that the intuitive strategy of {\em always} preferring the 
stationary point with the lowest SFT grand potential is unsystematic and 
therefore unacceptable.
The essence of the correct strategy, on the other hand, is to disregard,
as far as possible, those stationary points that (at a certain level of the
approximation) are trivially induced due to a partitioning of the reference
system into subsystems with fully decoupled degrees of freedom.

\begin{theacknowledgments}
The work is supported by the Deutsche Forschungsgemeinschaft within 
the Sonderforschungsbereich SFB 410 and the Forschergruppe FOR 538.
\end{theacknowledgments}

\bibliographystyle{aipproc}   

\end{document}